\documentclass[]{interact}

\usepackage[pdf]{pstricks}

\usepackage{epstopdf}
\usepackage{subfigure}

\usepackage[numbers,sort&compress,merge]{natbib}
\bibpunct[, ]{(}{)}{,}{n}{,}{,}

\theoremstyle{plain}

\theoremstyle{definition}

\theoremstyle{remark}

\begin{document}

\title{Development of a Digital Astronomical Intensity Interferometer: laboratory results with thermal light}

\author{
	\name{Nolan Matthews\thanks{CONTACT Author. Email: nolankmatthews@gmail.com}, David Kieda, and Stephan LeBohec}
	\affil{Department of Physics \& Astronomy, University of Utah}
}

\maketitle

\begin{abstract}
	We present measurements of the second order spatial coherence function of thermal light sources using Hanbury-Brown and Twiss interferometry with a digital correlator. We demonstrate that intensity fluctuations between orthogonal polarizations, or at detector separations greater than the spatial coherence length of the source, are uncorrelated but can be used to reduce systematic noise. The work performed here can readily be applied to existing and future Imaging Air-Cherenkov Telescopes used as star light collectors for Stellar Intensity Interferometry (SII) to measure spatial properties of astronomical objects.     
\end{abstract}


\section{INTRODUCTION}

Measurement of the correlation in intensity fluctuations of a light source gives access to the squared modulus of the complex degree of coherence. The pioneering experiments of Hanbury-Brown and Twiss (HBT) demonstrated that the modulus of the degree of coherence can be exploited to retrieve information about the morphology of astronomical objects \cite{HBT1957a}. This led to the construction of the Narrabri Stellar Intensity Interferometer (NSII) which was successfully used to measure 32 stellar diameters \cite{HB1974}. The advent of large Imaging Air-Cherenkov Telescope (IACT) arrays sparked a renewed interest in the stellar intensity interferometry (SII) technique \cite{SIIwIACT} and much recent work has been performed examining the ability to retrieve valuable astronomical observations with high angular resolution using a modern SII observatory. In particular, these include stellar imaging capabilities of IACT observatories \cite{paulthesis}, and laboratory setups using pseudo-thermal sources \cite{Dainis1,dainis2}. Other work on SII includes resolving the temporal coherence of a thermal source \cite{Tan1}, demonstration of the relative insensitivity to atmospheric turbulence \cite{Tan2}, investigations into temporal intensity interferometry using narrow-band emission lines from astrophysical sources \cite{Tan3}, and improving the obtainable SNR via the use of multi-channel intensity interferometry \cite{sascha}. 
\\

Since the intensity interferometry (II) technique measures the squared modulus of the degree of coherence, the phase information is lost, which complicates accurate image reconstruction. However, phase recovery is possible, given densely spaced coverage of the imaging plane, through both three-point correlations \cite{tripleproduct1} and Cauchy-Riemann algorithms \cite{paul2}. Practical implementations of modern SII have employed new technologies such as single photon detectors (SPDs) and high-speed digital data acquisition systems \cite{ASU2}. Initial measurements using some of these advancements were carried out using the Aquaeye+ and Iquaeye instruments which showed tentative measurements of coherence for a stellar source \cite{aquaeye}. 
\\

In this paper, we present new techniques for measuring the spatial coherence of a laboratory thermal source using high-speed photo-detectors and digital electronics. The modular nature of the detector and data acquisition system allows for straightforward integration with existing observatories. Parallel polarizations clearly demonstrate a photon bunching in time and space, whereas orthogonal polarizations eliminates coherence but reveals any additional correlation due to noise contamination. We show that correlation measurements in the orthogonal configuration, or when the detectors are separated at distances greater than the spatial coherence length of the source, can be used to correct for systematic noise due to spurious electronic correlations. 

\section{EXPERIMENTAL SETUP}

\begin{figure}
\centering
\includegraphics[width=\linewidth,height=0.6\linewidth]{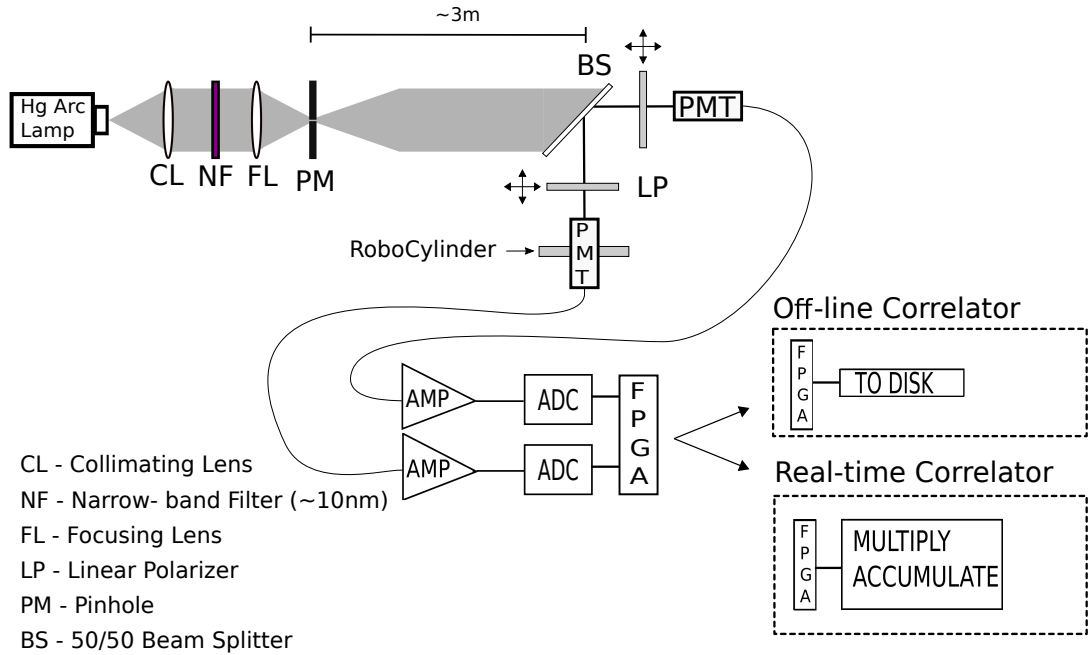}%
\caption{\label{fig:labsetup} Schematic of the laboratory setup}
\end{figure}

 A diagram of our laboratory system is shown in Fig. \ref{fig:labsetup}. Light from a mercury arc-lamp is collimated, passed through a 10$\,$nm narrow-band filter centered on the 435.8$\,$nm G spectral line of the Hg arc-lamp, and then refocused onto a spatial mask. The mask is either a single or double pinhole of various size configurations (typ. 200 - 300 micron diameter) simulating single and binary star systems. The output light passes through a long box (3$\,$m) and is split into two secondary beams via a 50/50 non-polarizing beam splitter. The light from each beam is then detected by super bi-alkali ($>$ 35\% Q.E.) high-speed photo-multiplier tubes (PMT). The PMTs used in the laboratory are the same as those currently employed on the cameras of an IACT observatory, VERITAS \cite{pmtupgrade}. The light collecting areas of the detectors are limited by a circular aperture of 5$\,$mm diameter. It is noted that the PMT aperture is of comparable size to the spatial coherence length of the source. This is done to increase the amount of light throughput into the detector such that the necessary integration time needed to reach a desired sensitivity level is reduced. The effects of large detector areas have already been described in other work \cite{janvida} and are taken into account here. The PMTs are also enclosed in a brass tube to shield them from unwanted electro-magnetic radiation. Linear optical polarizers may be placed in front of each PMT and can be individually rotated to select parallel or orthogonal polarization between the detectors. 

In order to sample different regions of the spatial coherence curve, one of the PMTs is mounted on a RoboCylinder linear actuator, whose position is controlled via LabVIEW software to high accuracy. The positioning is integrated into our data acquisition system allowing for automated measurements at varying positions. The output cables from the PMTs are fed into a low noise high-speed ($>$ 200$\,$MHz) FEMTO trans-impedance preamplifier. The resulting signal is sent through 10$\,$ft of double shielded cable (RG-223) and then continuously digitized by an analog-to-digital converter (ADC) at a rate of 250$\,$MS/s using an AC-coupled National Instruments (NI) FlexRIO adapter module (NI-5761). 

We have successfully employed two different types of digital correlators, off-line and real-time. In the off-line correlator, the digitized data from each channel is scaled, truncated to 8-bits, and merged into a single continuous data stream by a Virtex-5 FPGA (PXIe7965R). The data stream is then recorded to a high speed (700$\,$MB/s) 12TB RAID disk. A software routine using LabVIEW can later be used to retrieve intensity correlations between channels as a function of the digital time lag, typically up to $\pm$ 1$\,\mu$s in steps of 4$\,$ns. Due to the large number of samples, the data is read in blocks of 512 samples. The convolution theorem gives the correlation between two signals as the inverse Fourier transform of the product of the Fourier transforms of the signals. This is implemented by use of the NI Multi-core Analysis and Sparse Matrix toolkit (MASM) cross-correlation virtual instrument (VI), which optimizes the computation by utilizing separate computing cores. 

The largest drawback of performing the correlation off-line is the computation time needed to analyze the data. Data is read into a buffer using a single computing core, and then correlated using the remaining cores. Depending on the number of samples in each data block, it takes on the order of an hour of computation time for every minute of data recorded. This is mainly due to the time required to perform the correlation via the Fourier method for each data block in the NI cross-correlation VI. Since the correlation can be easily parallelized, this could be remedied by using a super-computer with many ($>$1000) processing cores. The NI controller used has only 4 cores limiting the maximum number of correlations performed at the same time. 

Some of the results presented herein were obtained by use of a real-time correlator using the Virtex-5 FPGA. In this implementation, the cross-correlation is computed using a multiply-accumulate algorithm with delay nodes to retrieve the correlation at various time lags with the FPGA clock set at 125$\,$MHz. The standard deviation of the entire correlogram excluding the zero time-delay bin is recorded and the time-stream of both channels are displayed on the LabVIEW front panel interface to allow visual inspection of the data. The FPGA clock for the algorithm is limited  to 125$\,$MHz (maximum of 250$\,$MHz). This timing restriction reduces the signal to noise ratio (SNR) by a factor of $\sqrt{2}$ in comparison to the off-line correlation. However, the correlations are retrieved in real-time allowing for immediate inspection of results and iterative tests of the laboratory setup. In the future, a compromise between the off-line and FPGA methods can be achieved by first streaming the data to disk, and then using the FPGA to perform the correlations on stored data.

\section{EXPERIMENTAL OBSERVABLES AND DATA REDUCTION}

\subsection{Review of II Measurements}

The correlation between AC-coupled amplified voltage signals, $J_1(t)$ and $J_2(t)$, from separated photo-detectors is,
\begin{equation}
c(\tau) = \frac{1}{T_0} \int_{0}^{T_0} J_1(t-\tau) J_2(t) dt
\end{equation}
where $T_0$ is the total integration time of the correlator, and $\tau$ is the time delay between channels. Hanbury-Brown and Twiss showed \cite{HBT1957b} that the correlation $\bar c(0)$ for a linearly polarized partially-coherent source of finite angular size could be written as,
\begin{equation}
\bar c(0) = 2e^2 A_1 A_2 \int_{0}^{\infty} |\Gamma (\nu,d)|^2 \, \alpha^2 (\nu) \, n^2 (\nu) d\nu \int_{0}^{\infty} |F(f)|^2 df
\end{equation}
where $A_1$ and $A_2$ are the light collection areas for each detector, $\alpha$ is the quantum efficiency, assumed to be the same for both channels, and $n$ is the spectral density of the source in units of photons sec$^{-1}$ Hz$^{-1}$ m$^{-2}$. $\Gamma$ is the coherence factor expected from the source and is dependent on the detector separation $d$. The term $F(f)$ represents the frequency response of the detectors and amplifiers. The optical bandwidth of the light, $\Delta \nu$ as set by filters in the optical system, is generally narrow enough that the quantum efficiency, spectral density, and coherence can be assumed as constant over the optical bandwidth. Additionally, for a rectangular bandpass the integral over the frequency response can be re-written as $\int_{0}^{\infty} |F(f)|^2 df = |F_{max}|^2 \, \Delta f $, where $|F_{max}|$ is the effective gain in a single channel (assuming identical channels), and $\Delta f$ is the electronic bandwidth of the correlator assuming that the gain is approximately constant over the electronic bandwidth. The correlation then becomes
\begin{equation}
\label{eqn:cbar}
\bar c(0) = 2e^2 A_1 A_2 \alpha^2 n^2 |F_{max}|^2 \Delta \nu \Delta f |\Gamma (d)|^2 
\end{equation}

The ability to detect the coherence of the source is limited due to shot noise fluctuations in each channel. Hanbury-Brown and Twiss showed that for identical channels the root mean square fluctuations in the correlator output due to shot noise is 
\begin{equation}
\label{eqn:rmsnoise}
\sigma = \sqrt{2} e^2 \alpha n \Delta \nu (A_1 A_2)^{\frac{1}{2}} |F_{max}|^2 ( \frac{\Delta f}{T_0} )^{\frac{1}{2}}. 
\end{equation}
To find the signal to noise ratio (SNR) we divide equation (\ref{eqn:cbar}) by equation (\ref{eqn:rmsnoise}) retrieving
\begin{equation}
\label{eqn:SNR}
{SNR} = \sqrt{2} (A_1 A_2)^{\frac{1}{2}} \, \alpha \, n \, | \Gamma (d)|^2  \sqrt{ \Delta f  T_0}.
\end{equation}

The above equation represents an idealistic form of the SNR. In this derivation, we assume point-like detectors that exhibit no dark current or after-pulsing. Furthermore, the above SNR does not include the contribution from stray light entering the detector, losses in the correlator, or pickup of additional noise in the data acquisition system. More complete treatments that include many of these additional considerations have already been performed \cite{HBT1957a,HBT1957b,janvida}.     
\\

In our experiment, we digitize the voltage such that time is discretized, $J(t) \rightarrow J(t_i)$, making the observed ADC reading,
\begin{equation}
K(t_i) = [ \frac{2^{n_b}}{V_r} J(t_i) ]
\end{equation}
where [ ] represents rounding the value to the nearest integer, $V_r$ is the voltage range of the digitizer, and $n_b$ is the number of resolution bits. The observed digital correlation is then,
\begin{equation}
c(t_k) = \frac{1}{T_0} \sum_{i=0}^{N} K_1(t_i - t_k) K_2(t_i) \Delta t .
\end{equation}
where $t_k$ is the discrete digital time delay, and $\Delta t$ is the sampling time of the ADC. 

\subsection{Correlated Noise Reduction - ON/OFF Analysis}

When operating at the large bandwidths required by an intensity interferometry system, there is often the undesired influence of spurious correlated noise degrading the spatial coherence measurement for a given source. Noise sources are varied, from electronic cross-talk between channels in the recording system, to Cherenkov light in the atmosphere due to gamma-rays when observing stars. In the laboratory, a persistent noise source is attributed to radio-frequency (RF) pickup. This RF signal is simultaneously detected in both electronic channels producing correlated noise. Regardless of the source, if the unwanted correlated noise is stable on operational timescales it can then be measured and removed. The exact behavior of each noise source on the correlated signal must be examined in a case-by-case basis. In this section, a general way to identify and reduce correlated noise by subtraction is presented. In our application, the temporal behavior of the correlated signal, or correllogram, is monitored over small time-lag windows ($<1\,\mu s$), throughout the integration process. In the laboratory total integration times are on the order of 5 - 20 minutes, but will be greater than one hour when observing stellar sources with telescopes.
\\

The measured correlation as a function of the time delay is
\begin{equation*}
c(\tau) = < K_1( t ) K_2( t+ \tau )>
\end{equation*}

Typically, the sampling time of the digitizer is much longer than the coherence time of the light. In this case, the correlation attributed to the spatial coherence of the source will only appear for the zero time-lag bin, $\tau$ = 0. For time-lags not equal to zero,  the correlation should be distributed randomly according to shot noise from photo-detection.
\\

Additional noise is then written as an additive term to the ADC reading recorded for each channel at the digitizer input, 
\begin{equation*}
K(t) = S(t) + N(t)
\end{equation*}
where $S(t)$ is the signal attributed to the source which includes both the wave and shot noise components, and $N(t)$ is the noise introduced into the system. In general, the noise term, $N(t)$, may result from a combination of several noise sources. The resulting correlation is then,
\begin{equation*}
\begin{aligned}
c(\tau) = &<S_1(t)S_2(t + \tau)> + <N_1(t)N_2(t + \tau)> \\ &+ <S_1(t)N_2(t)> 
+ <S_2(t+\tau)N_1(t)>
\end{aligned}  
\end{equation*}

The goal is then to identify and remove all above terms except for the correlation between $S_1$ and $S_2$. Now, it is necessary to consider at what stage in the measurement process is the noise introduced into the detection. For purely electronic noise which occurs after photo-detection, the cross-terms between the signal in one channel and noise in another, known as the cross-talk, is ignored. In the laboratory, we observe that the cross-talk between the channels is negligible compared to the signal and noise correlations. The measured correlation can then be written as
\begin{equation*}
c(\tau) = <S_1(t) S_2(t+\tau) > + <N_1(t) N_2(t+\tau)>
\end{equation*}
where only the noise not correlated to the signal itself was kept. The correlated noise appears as a purely additive term to the overall correlation. 
\\
 
To remove the correlated noise, we perform a background measurement of the correlation which does not contain the signal attributed to the spatial coherence, but, includes the noise contribution at the same level as in the desired correlation measurement. The final measurement is obtained as the residual between the ON observation, where source coherence is expected, and the OFF observation. A straightforward way to obtain OFF data in the laboratory is to measure the correlation for detector separations large enough for the contribution due to the coherence of the source to be negligible. This makes the observed correlation
\begin{equation*}
\begin{aligned}
c_{F} (\tau,T,d) = & < S_1(t) S_2(d_{on},t+\tau) > - <S_1(t+T) S_2(d_{off},t+T+\tau) > \\ &+ <N_1(t) N_2(d_{on},t+\tau)> - <N_1(t+T) N_2(d_{off},t+T+\tau)>
\end{aligned}
\end{equation*}
where $T$ is the time difference between recordings of the ON and OFF runs, and $d_{on}$ and $d_{off}$ are the detector separations in the ON and OFF configurations. Given a circular source with angular diameter $\theta_d$, the detector separation for the background correlation must be greater than  $1.22\lambda/\theta_d$ so that the coherence from the source is very small. Ideally, the noise sources do not significantly change between ON and OFF runs such that the residual between noise correlations tends to zero, leaving only the difference in signal correlations. In order to alleviate for the slow changes in noise level between ON and OFF observations we tend to proceed with relatively rapid observation cycles of no more than a few minutes period. 
\\
\begin{figure}[t]
	\centering
	\includegraphics[width=\linewidth]{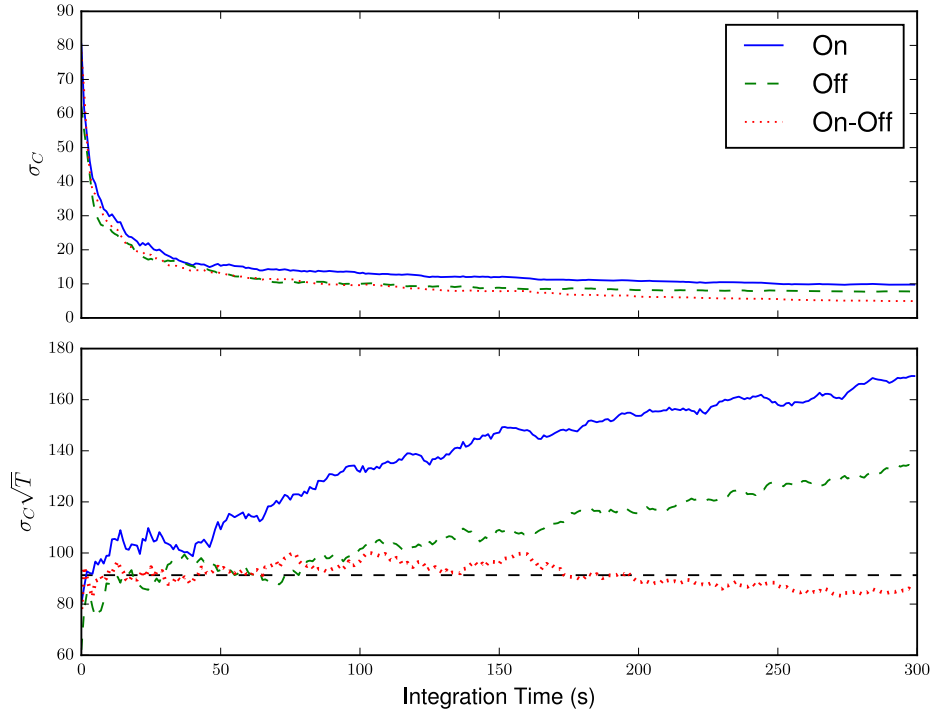}
	\caption{The top panel shows the standard deviation of the correllogram excluding the zero time-lag bin as a function of the total integration time. In the bottom panel the same data is shown but multiplied by $\sqrt{T_0}$. The horizontal black dashed line shows the mean of the ON-OFF analysis. For both the ON and OFF runs, the presence of spurious correlations causes the R.M.S. trend to deviate from the expected $\frac{1}{\sqrt{T_0}}$. In the case of the ON-OFF analysis the R.M.S. tends to follow the expected trend.}
	\label{fig:stdanalysis}
\end{figure}

To ensure that the noise subtractions are properly performed, the R.M.S. distribution over the entire correlogram excluding the zero time-lag is monitored against the expected trend of $\frac{1}{\sqrt{T_0}}$, where $T_0$ is the total integration time. Initially, the shot noise component will dominate the R.M.S, but as the integration of the correlator proceeds, low-level noise correlations may be detected which sets a limit on the minimum detectable R.M.S. When the noise correlation is significant, the R.M.S. trend will deviate from $\frac{1}{\sqrt{T_0}}$. For proper noise subtraction the residual between the ON and OFF correlations should follow the $\frac{1}{\sqrt{T_0}}$ trend. Figure \ref{fig:stdanalysis} shows a typical result in the laboratory for the R.M.S. trend. An ON and then OFF run of 5 minutes were taken sequentially. The integrated correlation for ON, OFF, and ON-OFF was recorded every second and the R.M.S. was calculated for each measurement over the entirety of the integration time. The bottom panel displays the R.M.S. multiplied by the $\sqrt{T_0}$ such that the expected value should fluctuate about a constant. For both the ON and OFF runs it begins to deviate from the expected trend after only 50 seconds of integration (when the noise is detected). However, the residual between the ON and OFF runs appears to be more stable suggesting that the noise subtraction is being performed properly. Here, the normalization for the R.M.S. trend for ON and OFF runs is different which we attribute to varying levels in the light intensity and also the noise.
\\

The residual observation can also be performed using parallel (ON) and orthogonal polarization (OFF) configurations between the detectors. Light between orthogonal polarizations should show no coherence and thus can be used as a background, or OFF observation. This method provides an additional benefit since both parallel and orthogonal configurations can be observed simultaneously for a single detector separation.

\section{RESULTS}

\subsection{Validation of ON/OFF analysis}
\begin{figure}
	\centering
	\includegraphics[width=\linewidth,height=0.6\linewidth]{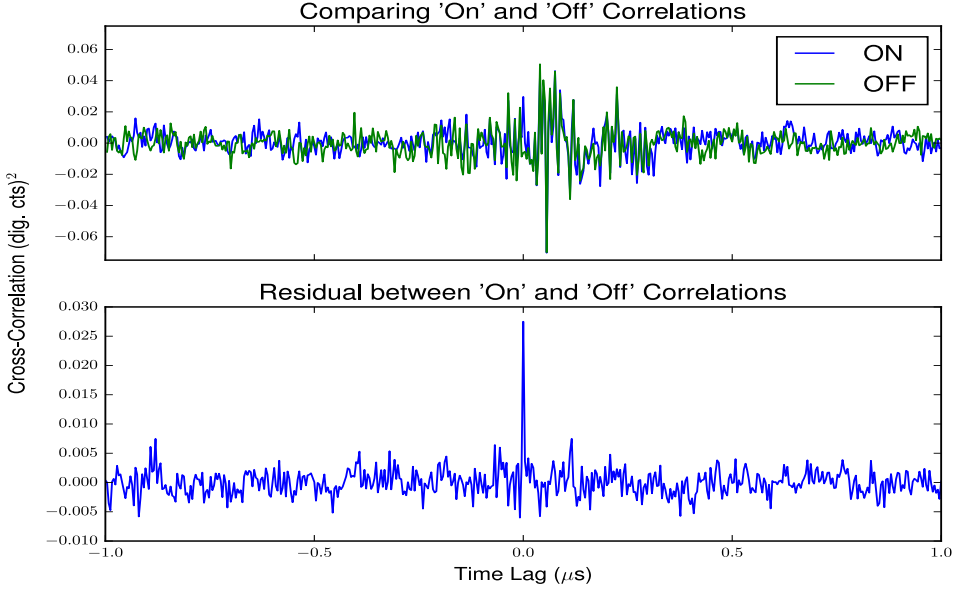}
	\caption{In the top panel sequential ON and OFF measurements of the correlogram over an integration time of 30 seconds each are overlaid. For small time lags ($\tau < 500\,$ns) the scatter between the measured correlation for different time lags increases significantly due to the presence of correlated noise. In the bottom panel, the residual between a total of 10 minutes each of ON and OFF data (comprised of 30 second sequential runs alternating between ON and OFF) is shown which reveals the zero time-lag correlation emanating from the spatial coherence of the source.}
	\label{fig:onoff_resid}
\end{figure}

The ON/OFF analysis was validated in the laboratory with the experimental setup shown in Fig. \ref{fig:labsetup} using the off-line correlator and without the use of polarizing filters. Fig. \ref{fig:onoff_resid} displays the correlogram both before and after obtaining the residual between ON and OFF runs. The ON region was chosen at zero baseline separation, and the OFF at a separation of 10 mm. Given the expected angular size of the source the first zero of the coherence function is reached at approximately 5.5$\,$mm. The subtraction of spurious noise reveals the coherence of the source at the zero time-lag bin.

\subsection{Spatial Coherence Measurement}

\begin{figure}[h]
	\centering
	\includegraphics[width=0.9\linewidth]{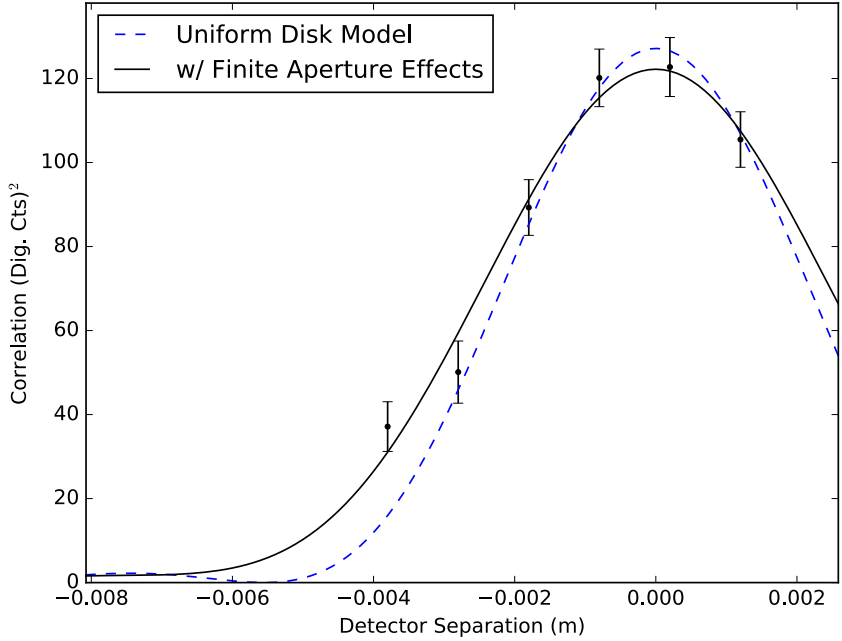}
	\caption{The above image shows the measured correlation from the FPGA correlator as a function of the detector separation. The dotted blue line is a fit to the data assuming a uniform disk model for the light source. The wavelength and diameter of the source is held constant but the normalization and center can vary. The black line is a similar fit, but includes the effects of the extended detector aperture.}
	\label{fig:spatialcoherence}
\end{figure}

To measure the spatial coherence a LabVIEW routine was developed which integrated the actuator movement with the data acquisition. The source consisted of a circular pinhole of approximately 300 micron diameter at a distance of 3.15$\,$m with a central wavelength of $\lambda =$ 435$\,$nm. Correlations were recorded at each position in 5 minute segments for both an ON and subsequent OFF run. A total of 6 ON positions each separated by 1mm were recorded about the zero baseline position. After each 5 minute integration, the mean of the correlogram, excluding the value at the zero time-lag bin, was subtracted from the entire correlogram. The residual between ON and OFF runs was then calculated. This process was repeated 4 times yielding a total integration time of 20 minutes at each position. 
\\

The result of this procedure is shown in Fig. \ref{fig:spatialcoherence}. The uncertainty in each measurement was determined by the RMS scatter for time lags away from zero. The dashed line represents a fit to the data by modeling the source as a uniform disk with fixed wavelength and angular diameter. The zero baseline (or center position) and normalization are left as free parameters and determined by the fit. The solid line includes the effects of the extended detector size \cite{janvida}. To include these effects in the fit, an initial model is generated by convolving the detector areas with the expected normalized spatial coherence. The resulting model was interpolated and then fit to the data in a similar manner as the initial fit without the detector size effects.
\\

A reduced $\chi ^2$ test was performed between the uniform disk model with detector size effects and the measured spatial correlation finding $\chi ^2 /\nu$ = 0.83 suggesting agreement between the data and model. However, there are several considerations for the source that are not taken into account here. Examination of the pinhole under a scanning electron microscope revealed irregularities in the diameter on the order of 5-10$\,\%$. Additionally, the angular brightness distribution may not be constant over the area of the pinhole, making the uniform disk model assumption not fully valid. 

\subsection{Correlation between orthogonal and parallel polarized thermal light}

\begin{figure}
	\centering
	\includegraphics[width=0.8\linewidth]{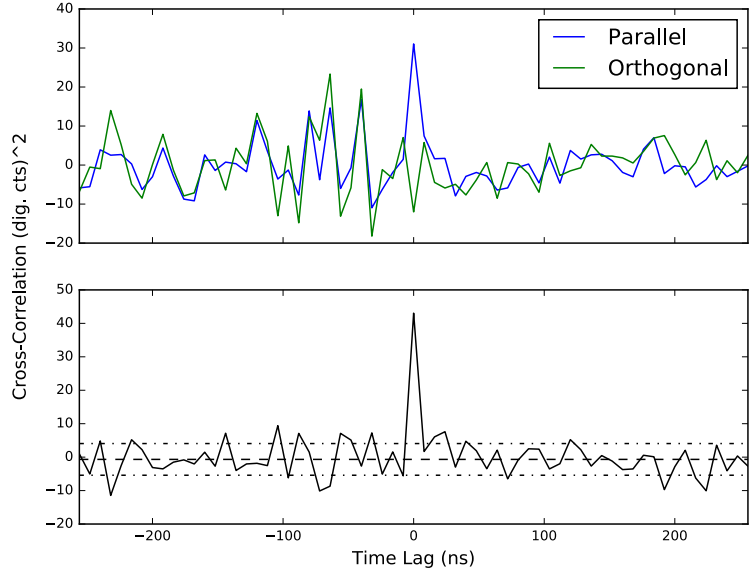}
	\caption{Correlogram for the polarization tests performed in the laboratory. The top panel shows the result for both the parallel and orthogonal configuration and the bottom panel shows the residual along with $\pm 1 \sigma$ indicators.}
	\label{fig:polarizationResid}
\end{figure}

The experiment was setup using the polarizing filters in a parallel configuration in front of each detector. Real-time FPGA correlations for minimal detector separation were recorded for a period of 5 minutes. The filter was manually rotated by 90 degrees to select the orthogonal configuration and the correlation measurement was repeated. The results are shown in Fig. \ref{fig:polarizationResid} which shows the correlogram both before and after the application of the ON/OFF subtraction. The noise subtraction between parallel and orthogonal polarizations offered an improvement of 59$\%$ of the SNR over the parallel configuration measurements.

\section{SUMMARY AND OUTLOOK}

The second order coherence function for simulated stars using a thermal light source were measured with a digital correlator. An ON/OFF analysis routine was developed allowing for removal of systematic spurious correlations due to unwanted noise pickup. The routine consisted of either physically separating the detectors so that the coherence from the source is negligible or using orthogonal polarizations in order to measure a background. The main application of this work is towards a modern SII array using IACT arrays to observe stars. The system will be integrated into the StarBase-Utah \cite{starbase} observatory over the Summer of 2017 for initial tests to verify operation on actual astronomical telescopes, and then scaled to the VERITAS array in the Fall 2017 and Winter 2017-2018. 
\\

Our group is actively working to improve some of the features presented in this paper. At the time when these results were obtained the maximum bandwidth of the FPGA correlator was limited due to timing restrictions. However, a modified algorithm was developed allowing for optimal operation of the FPGA correlator. Additionally, instead of performing the correlations in real-time, the data can be first streamed to disk, where the correlations are performed post-digitization. A benefit of streaming data to disk is that an arbitrary number of channels can be correlated with computation time as the only limitation. This opens the possibility for correlations between selected polarization modes as well as multiple spectral channels. The obtainable SNR is then improved by the square root of the number of additional channels. Also, proper normalization of the correlation between runs for varying light levels has yet to be demonstrated. The true normalization depends on a number of factors, primarily the light intensity and gain variations. Within small integration times in the laboratory the light intensity and gain can be expected to be constant, however, over hour-long time scales as needed for stellar observations, these changing parameters need to be accounted for. The normalization of the correlation for varying light intensity and gain fluctuations has already been studied by Hanbury-Brown and Twiss \cite{hbtbook}.
\\

For the integration into IACT telescopes there are several tasks to be demonstrated. First, to use the ON/OFF analysis it is necessary to measure the orthogonal and parallel polarization of light simultaneously to remove the lasting effect of any transient noise sources as well as reducing total data collection time. A straightforward implementation of this is to use a polarized beam splitter to separate the orthogonal polarizations. Each telescope will have its own data acquisition hardware with the data brought together to a central processing unit post-digitization. This requires synchronization of the ADC modules to sub nanosecond precision which already can be accomplished using fiber optics and external clocks \cite{whiterabbit}. In the laboratory we have already achieved synchronization for closely spaced ($<$ 1m) but physically separated data acquisition modules using a central timing unit and coaxial cable connections. However, this capability still needs to be demonstrated over large ($>$ 100m) distances.  
\\

We have successfully measured the coherence of a thermal blackbody source in both time and space using a digital correlator. Small modifications to the current experimental setup allow interferometric capabilities on large arrays of IACTs at very modest costs and are currently being pursued.
\\
\section*{ACKNOWLEDGEMENTS}
This is an author's Accepted Manuscript of an article published by Taylor \& Francis in the Journal of Modern Optics on August 10, 2017, available online at: http://www.tandfonline.com/10.1080/09500340.2017.1360958."
\\

The authors would like to dedicate this work to Micah Kohutek for his work in the laboratory with regrets that he did not get to see this recent progress. We would like to also acknowledge Udara Abeysekara for helpful discussions and assistance in the setup of the linear actuator. The authors gratefully acknowledge for this work from the University of Utah and from National Science Foundation grants PHY151050 and PHY0960242.

\newpage
\bibliographystyle{tfp}
\bibliography{SIILabPaper_JMO_edit}

\begin{thebibliography}{10}
\providecommand{\url}[1]{\normalfont{#1}}
\providecommand{\urlprefix}{}

\bibitem{HBT1957a}
{Hanbury Brown}, R.; {Twiss}, R.Q. {Interferometry of the Intensity
  Fluctuations in Light. I. Basic Theory: The Correlation between Photons in
  Coherent Beams of Radiation}, \emph{Proceedings of the Royal Society of
  London Series A}  \textbf{1957}, \emph{242}, 300--324.

\bibitem{HB1974}
Hanbury~Brown, R.; Davis, J.; Allen, L.R. The Angular Diameters of 32 Stars,
  \emph{Monthly Notices of the Royal Astronomical Society}  \textbf{1974},
  \emph{167}~(1), 121. \urlprefix\url{+
  http://dx.doi.org/10.1093/mnras/167.1.121}.

\bibitem{SIIwIACT}
LeBohec, S.; Holder, J. Optical Intensity Interferometry with Atmospheric
  Cerenkov Telescope Arrays, \emph{The Astrophysical Journal}  \textbf{2006},
  \emph{649}~(1), 399.
  \urlprefix\url{http://stacks.iop.org/0004-637X/649/i=1/a=399}.

\bibitem{paulthesis}
Nunez, P.D. Towards optical intensity interferometry for high angular
  resolution stellar astrophysics. Ph.D. Thesis, The University of Utah, 2012.

\bibitem{Dainis1}
Dravins, D.; Lagadec, T.; Nunez, P.D. Long-baseline optical intensity
  interferometry, \emph{A \& A}  \textbf{2015}, \emph{580}.

\bibitem{dainis2}
{Dravins}, D.; {Lagadec}, T.; {Nu{\~n}ez}, P.D. {Optical aperture synthesis
  with electronically connected telescopes}, \emph{Nature Communications}
  \textbf{2015}, \emph{6}, 6852.

\bibitem{Tan1}
Tan, P.K.; Yeo, G.H.; Poh, H.S.; Chan, A.H.; Kurtsiefer, C. Measuring Temporal
  Photon Bunching in Blackbody Radiation, \emph{The Astrophysical Journal
  Letters}  \textbf{2014}, \emph{789}~(1), L10.
  \urlprefix\url{http://stacks.iop.org/2041-8205/789/i=1/a=L10}.

\bibitem{Tan2}
{Tan}, P.K.; {Chan}, A.H.; {Kurtsiefer}, C. {Optical intensity interferometry
  through atmospheric turbulence}, \emph{MNRAS}  \textbf{2016}, \emph{457},
  4291--4295.

\bibitem{Tan3}
{Tan}, P.K.; {Kurtsiefer}, C. {Temporal intensity interferometry for
  characterization of very narrow spectral lines}, \emph{MNRAS}  \textbf{2017},
  \emph{469}, 1617--1621.

\bibitem{sascha}
{Trippe}, S.; {Kim}, J.Y.; {Lee}, B.; {Choi}, C.; {Oh}, J.; {Lee}, T.; {Yoon},
  S.C.; {Im}, M.; {Park}, Y.S. {Optical Multi-Channel Intensity Interferometry
  - Or: How to Resolve O-Stars in the Magellanic Clouds}, \emph{Journal of
  Korean Astronomical Society}  \textbf{2014}, \emph{47}, 235--253.

\bibitem{tripleproduct1}
{Wentz}, T.; {Saha}, P. {Feasibility of observing Hanbury Brown and Twiss
  phase}, \emph{MNRAS}  \textbf{2015}, \emph{446}, 2065--2072.

\bibitem{paul2}
{Nu{\~n}ez}, P.D.; {Holmes}, R.; {Kieda}, D.; {Rou}, J.; {LeBohec}, S. {Imaging
  submilliarcsecond stellar features with intensity interferometry using air
  Cherenkov telescope arrays}, \emph{MNRAS}  \textbf{2012}, \emph{424},
  1006--1011.

\bibitem{ASU2}
{Pilyavsky}, G.; {Mauskopf}, P.; {Smith}, N.; {Schroeder}, E.; {Sinclair}, A.;
  {van Belle}, G.T.; {Hinkel}, N.; {Scowen}, P. {Single-Photon Intensity
  Interferometry (SPIIFy): utilizing available telescopes}, \emph{MNRAS}
  \textbf{2017}, \emph{467}, 3048--3055.

\bibitem{aquaeye}
{Zampieri}, L.; {Naletto}, G.; {Barbieri}, C.; {Barbieri}, M.; {Verroi}, E.;
  {Umbriaco}, G.; {Favazza}, P.; {Lessio}, L.; {Farisato}, G. {Intensity
  interferometry with Aqueye+ and Iqueye in Asiago}, In \emph{Optical and
  Infrared Interferometry and Imaging V}, Aug, 2016; Proc. SPIE, Vol. 9907, p
  99070N.

\bibitem{pmtupgrade}
Otte, A.N. The Upgrade of VERITAS with High Efficiency Photomultipliers, In
  \emph{32nd International Cosmic Ray Conference}, 2011.

\bibitem{janvida}
{Rou}, J.; {Nu{\~n}ez}, P.D.; {Kieda}, D.; {LeBohec}, S. {Monte Carlo
  simulation of stellar intensity interferometry}, \emph{MNRAS}  \textbf{2013},
  \emph{430}, 3187--3195.

\bibitem{HBT1957b}
{Hanbury Brown}, R.; Twiss, R.Q. Interferometry of the Intensity Fluctuations
  in Light II. An Experimental Test of the Theory for Partially Coherent Light,
  \emph{Proceedings of the Royal Society of London A: Mathematical, Physical
  and Engineering Sciences}  \textbf{1958}, \emph{243}~(1234), 291--319.
  \urlprefix\url{http://rspa.royalsocietypublishing.org/content/243/1234/291}.

\bibitem{starbase}
{LeBohec}, S.; {Adams}, B.; {Bond}, I.; {Bradbury}, S.; {Dravins}, D.;
  {Jensen}, H.; {Kieda}, D.B.; {Kress}, D.; {Munford}, E.; {Nu{\~n}ez}, P.D.;
  et~al. {Stellar intensity interferometry: experimental steps toward
  long-baseline observations}, In \emph{Optical and Infrared Interferometry
  II}, July, 2010; Proc. SPIE, Vol. 7734, p 77341D.

\bibitem{hbtbook}
{Hanbury Brown}, R. \emph{The Intensity Interferometer. Its applications to
  astronomy}; Taylor \& Francis, 1974.

\bibitem{whiterabbit}
{Serrano}, J.; {Alvarez}, P.; {Cattin}, M.; {Cota}, E.; {P. M. J. H. Lewis};
  T., W.; {et. al,}. {The White Rabbit Project in Proceedings of ICALEPCS
  TUC004}, In , KOBE, Japan, 2009.

\end{thebibliography}

\end{document}